\documentclass[manuscript]{aastex}
\usepackage{natbib}
\usepackage{color}
\usepackage{ulem}
\usepackage{xcolor}
\shorttitle{}
\shortauthors{Hsieh et al.}

\begin{document}

\title{The Fossil Nuclear Outflow in the Central 30 pc of the Galactic Center}

\author{
        Hsieh, Pei-Ying\altaffilmark{1,2}, Ho, Paul T. P. \altaffilmark{1,3}, Hwang, Chorng-Yuan\altaffilmark{2}, Shimajiri, Yoshito\altaffilmark{4,5}, Matsushita, Satoki \altaffilmark{1}, Koch, Patrick M. \altaffilmark{1}, \& Iono, Daisuke\altaffilmark{5,6}
\\pyhsieh@asiaa.sinica.edu.tw}

\affil{$^1$ Academia Sinica Institute of Astronomy and
       Astrophysics, P.O. Box 23-141, Taipei 10617, Taiwan, R.O.C.}
\affil{$^2$ Institute of Astronomy, National Central University,
       No.300, Jhongda Rd., Jhongli City, Taoyuan County 32001, Taiwan, R.O.C.}
\affil{$^3$ East Asian Observatory, 660 N. Aohoku Place, University Park, Hilo, Hawaii 96720, U.S.A.}
\affil{$^4$ Laboratoire AIM, CEA/DRF–CNRS–Universit\'{e} Paris Diderot, IRFU/Service d'Astrophysique, C.E. Saclay, Orme des Merisiers, 91191 Gif-sur-Yvette, France}
\affil{$^5$ National Astronomical Observatory of Japan, 2-21-1 Osawa, Mitaka, Tokyo 181-8588, Japan} 
\affil{$^6$ The Graduate University for Advanced Studies (SOKENDAI), 2-21-1 Osawa, Mitaka, Tokyo 181-8588, Japan}

\begin{abstract}

We report a new 1-pc (30$\arcsec$) resolution CS($J=2-1$) line map of the central 30 pc of the Galactic Center (GC), made with the Nobeyama 45m telescope. 
We revisit our previous study of the extraplanar feature called polar arc (PA), which is a molecular cloud located above SgrA* with a velocity gradient perpendicular to the Galactic plane. We find that the PA can be traced back to the Galactic disk. This provides clues of the launching point of the PA , roughly $6\times10^{6}$ years ago. Implications of the dynamical time scale of the PA might be related to the Galactic Center Lobe (GCL) at parsec scale.
Our results suggest that in the central 30 pc of the GC, the feedback from past explosions could alter the orbital path of the molecular gas down to the central tenth of parsec.
In the follow-up work of our new CS($J=2-1$) map, we also find that near the systemic velocity, the molecular gas shows an extraplanar hourglass-shaped feature (HG-feature) with a size of $\sim$13 pc. 
The latitude-velocity diagrams show that the eastern edge of the HG-feature is associated with an expanding bubble B1, $\sim$7 pc away from SgrA*. The dynamical time scale of this bubble is $\sim3\times10^{5}$ years. This bubble is interacting with the 50 km s$^{-1}$ cloud. Part of the molecular gas from the 50 km s$^{-1}$ cloud was swept away by the bubble to $b=-0.2\degr$.  
The western edge of the HG-feature seems to be the molecular gas entrained from the 20 km s$^{-1}$ cloud towards the north of the Galactic disk. Our results suggest a fossil explosion in the central 30 pc of the GC a few 10$^{5}$ years ago.

\end{abstract}


\section{INTRODUCTION}\label{sect-intro}

Our Galactic Center (GC) is the nearest nucleus of a galaxy ($d$=8.5 kpc) \citep{reid93,ghez08,reid14}. It is the best target to study detailed structures and dynamics in a circumnuclear environment at sub-pc scale, which cannot be easily done in external galaxies with ground-based telescopes. However, observations of the GC suffer from our edge-on vantage point. Optical and near infrared (IR) emission suffers from large amounts of extinction. The V-band extinction ($A_{\rm v}$)
varies from 20 to 50 mag with a median value of 31.1 mag \citep{scoville03}.
Dust is transparent at radio wavelengths. However, at millimeter wavelengths, the more abundant molecular lines (e.g., CO, J = 1--0) become optically thick and suffer from foreground absorption or self-absorption in the direction towards the GC \citep{guesten87,wright01,chris05}. High-excitation molecular lines, or high-density tracers are less affected by the foreground/ambient cold gas \citep{maria,tsuboi99,jackson93,mcgary01,mcgary02,herrnstein02,herrnstein05}.

Fruitful studies were already carried out with low-excitation molecular lines and with low angular resolutions \citep[e.g.][]{sco72,burton83,bally87,bally88,burton92,tsuboi99}.
Our focus is on the complex activities and physical conditions in the nuclear region probed by the dense molecular gas.
We conducted wide-field single-dish observations of the central 30 pc of the GC with multiple transitions of the CS molecule \citep[][hereafter paper I]{hsieh15}.
In paper I, we reported a new feature called the connecting ridge (CR) which was detected with the CS($J=4-3$) line. The CR has a velocity gradient perpendicular to the disk rotation.
It is physically associated with the extraplanar polar arc (PA)  \citep{bally88,henshaw16}.
The PA extends from north of the SgrA* region at a 40$\degr$ angle and shows a large velocity gradient from $(l,b,V_{\rm sys}) = (0\degr,0\fdg05,70~\rm km~s^{-1}$) to $(0\fdg2,0\fdg25,140~\rm km~s^{-1})$.  Below $V_{\rm sys}$ of 70 km s$^{-1}$, the PA lies close to the Galactic plane and becomes confused with the molecular clouds in the SgrA* region.
In paper I, we find that the kinematic and spatial structures connect the Galactic disk, the CR, and the PA. These results suggest that the molecular gas might be lifted out of the Galactic plane. We, thus, proposed the idea of a molecular outflow in the central 30 pc of the GC and suggested that the PA is pushed away, possibly by the energy of 8-80 supernovae explosions.
The importance of Galactic outflows is that they may be the primary mechanism to recycle metals and to deposit them into the intergalactic medium \citep{veilleux05}.
In starburst galaxies, vast amounts of stellar winds
from massive stars as well as supernovae explosions generate a
huge amount of energy and high pressure to create high-velocity
galactic winds, which interact with and sweep up the ambient
gas. 
The GC is the nearest nucleus, and therefore, it is the best target to resolve the structure, kinematics, and physical conditions of nuclear outflows. The presence of atomic and molecular gas allows us to measure the outflow of neutral material, its impact, and transfer of energy and momentum 
to the surrounding interstellar medium (ISM).
In this paper, we continue to investigate the possibility of the outflow nature in the GC with our new CS($J=2-1$) data.

\section{OBSERVATIONS AND DATA REDUCTION}\label{sect-obs}

The central molecular zone (CMZ) of the Milky Way was observed in the CS(J = 2--1; 97.98093 GHz) line with the 45m telescope of the Nobeyama Radio Observatory (NRO)\footnote{Nobeyama Radio Observatory is a branch of the National Astronomical Observatory of Japan, National Institutes of Natural Sciences.}
in May 2012. The full width at half maximum (FWHM) of the beam size at this frequency was $\sim17\arcsec$.
We used the 25 elements of the focal plane receiver
(BEam Array Receiver System: BEARS) \citep{sunada00} to
observe the central 108$\arcmin\times27\arcmin$ ($l\times b$) area
of the CMZ in the On-The-Fly (OTF) mapping mode \citep{sawada08}
with position-switching. 
As the backend spectrometer
we used digital auto-correlators
with a bandwidth of 512 MHz 
with a spectral resolution of 500 kHz with 1024 channels. The total bandwidth was $\sim$1740 km s$^{-1}$ with a resolution of 1.3 km s$^{-1}$ in the raw data.

The mapping area was arranged into four sub-regions, each covering 27$\arcmin\times27\arcmin$. 
We sampled an emission-free reference (OFF)
position at ($l,b$) = (1.$\degr$,--0.7$\degr$) for every two 35-second
on-source scans along each sub-region. The sampling interval along the scan
rows was 5\farcs1, and the separation between the scan rows
was 7\farcs5, which corresponds to roughly
1/3 and 1/2 of the beam. We scanned along the $l$ and $b$ directions in order to minimize scanning artefacts with the basket-weave method \citep{basket}. These artefacts originate
likely from pointing errors, time variations of the system temperature $T_{\rm sys}$, insufficient sampling grids, and the non-uniform beam separations
of the BEARS elements \citep{sawada08}.

The pointing of the antenna was corrected by measuring the SiO maser VX Sgr every hour with the SIS receiver S40 in the 40-GHz band.
The Double Side Band (DSB) $T_{\rm sys}$ varied from 300 to 700 K during our observations. We used the standard chopper wheel
method to calibrate the output signal into the antenna temperature
$T_{\rm A}^{*}$, which was corrected for atmospheric attenuation. Since the original
$T_{\rm sys}$ measured by BEARS were DSB
$T_{\rm sys}$, we needed to measure the relative gains of the two sidebands. We used the single beam receiver S100,
equipped with a Single Side Band (SSB) filter to observe a calibrator.
Every beam of BEARS was then scaled to
convert from $T_{\rm A}^{*}$(DSB) to $T_{\rm A}^{*}$(SSB).
The main beam efficiency $\eta_{\rm MB}$ was $\sim43\%$.
In the maps presented here we are using the $T_{\rm A}^{*}$ (SSB) scale.

We used the NOSTAR package \citep{sawada08} to reduce our
OTF data. The baselines were subtracted with linear or higher-order
polynomial functions, and bad scans were flagged. During the
observations, one
of the receivers (A09) had abnormally high $T_{\rm sys}$. Therefore,  the data of A09 were flagged, and we only used 24 receivers.
The resulting gridding size of the map is 7.5$\arcsec$, with an effective final resolution of 21$\arcsec$ with bessel-gaussian convolution and an rms noise of 0.1 K ($T_{\rm A}^{*}$) at a velocity resolution of 2.5 km s$^{-1}$.
In this paper, we present the CS($J=2-1$) line data for the central 30$\arcmin$ (70 pc) region of Milky Way. The entire CS($J=2-1$) data of the CMZ will be presented in another paper.

\section{The Central 30 pc of the Milky Way}
\subsection{CS($J=2-1$) Line Map}\label{cs21-line-map}

\begin{figure}[hp]
\begin{center}
\epsscale{0.5}
\includegraphics[angle=0,scale=0.4]{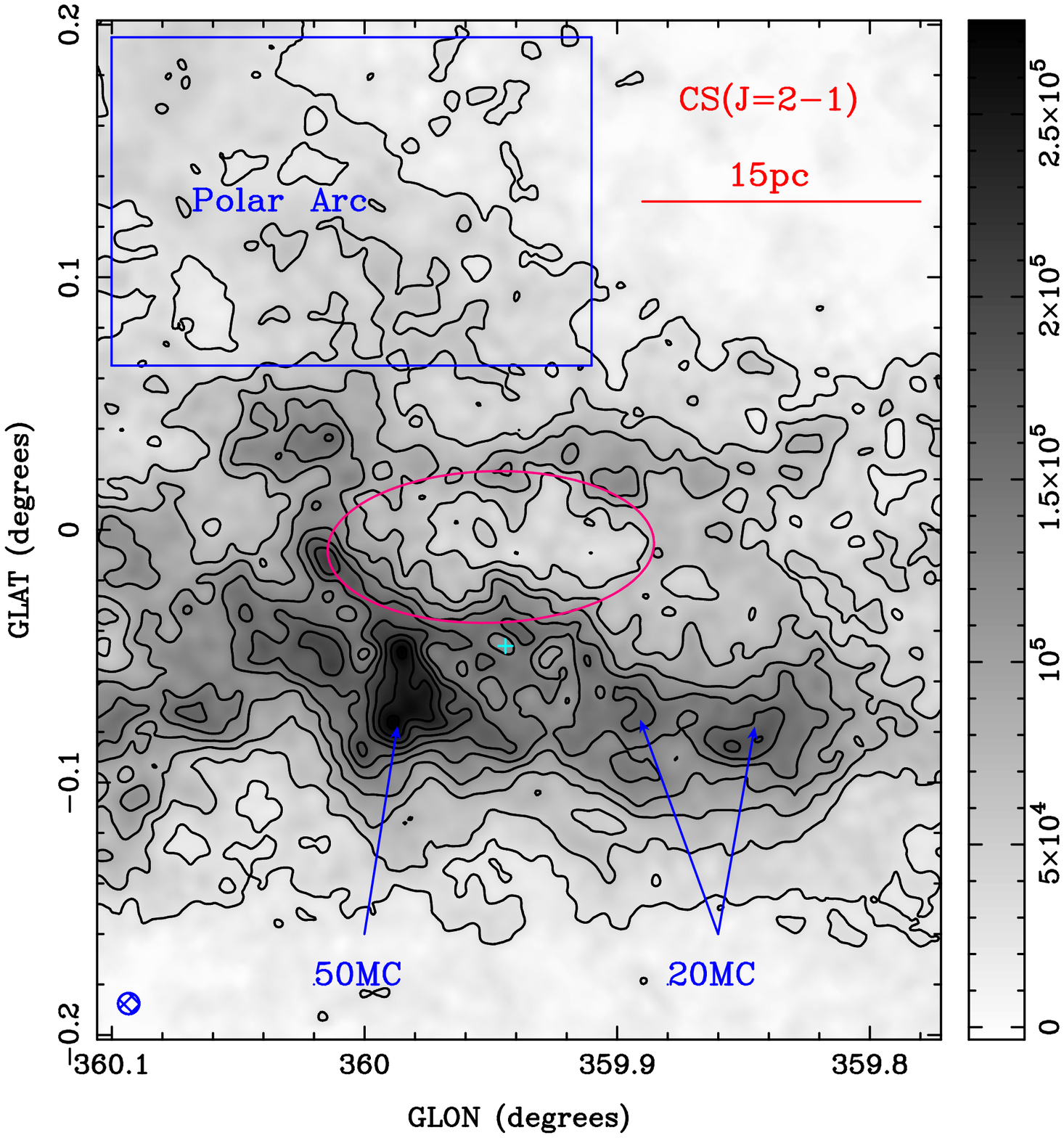} 
\includegraphics[angle=0,scale=0.4]{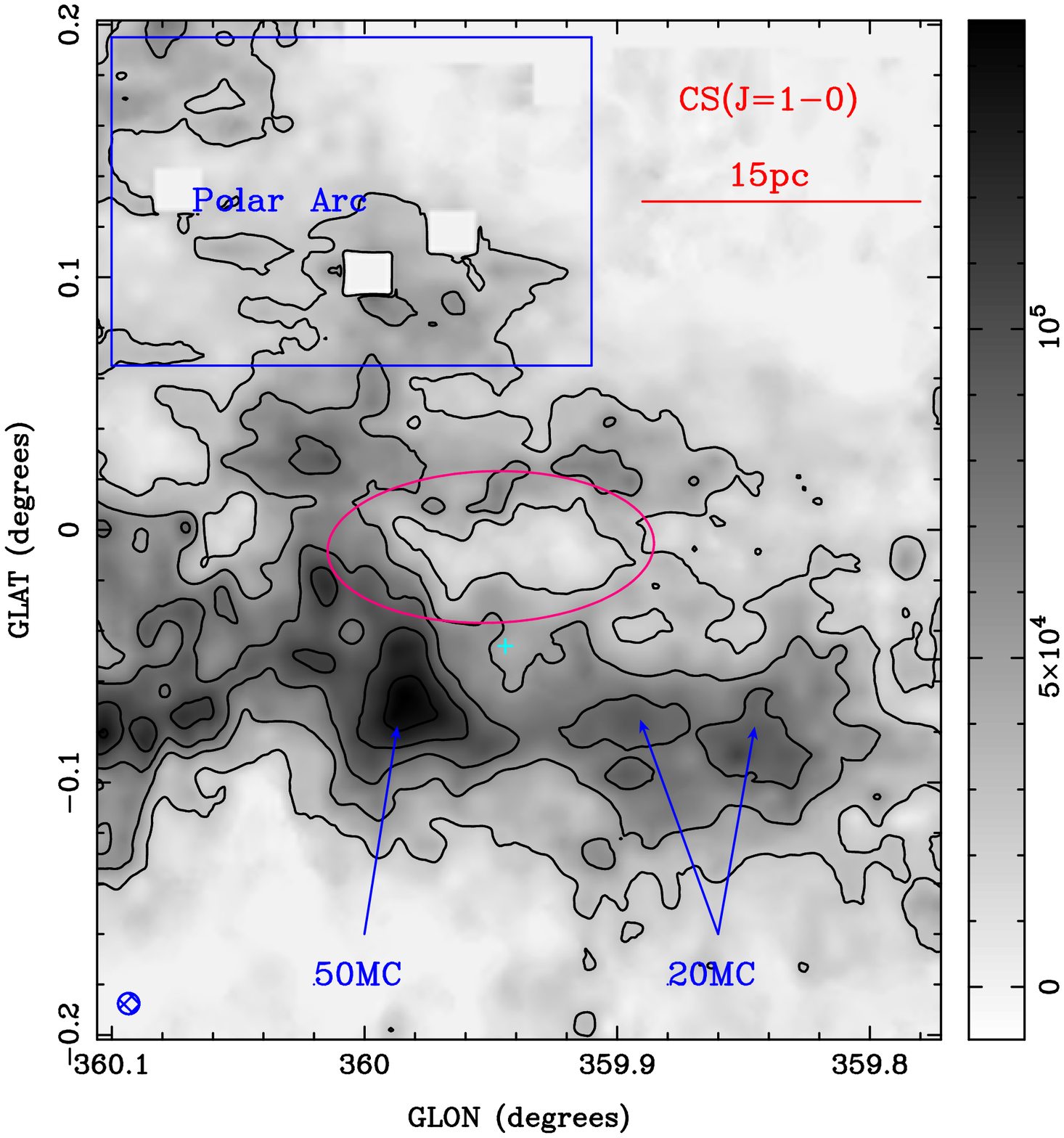} 
\caption[]{\small
Top: CS($J=2-1$) emission integrated from $-197.5$ km s$^{-1}$ to 197.5 km s$^{-1}$ ($T_{\rm A*}$ in unit of K m s$^{-1}$). Bottom: CS($J=1-0$) line emission \citep{tsuboi99} integrated from $-197.5$ km s$^{-1}$ to 197.5 km s$^{-1}$.  The beam size of the CS($J=2-1$) and CS($J=1-0$) line maps is 30$\arcsec$ (blue circles in lower left corners). The contours are 5, 10, 15, ..., 50$\times$5400 K m s$^{-1}$.  The positions of SgrA* (cyan cross), the 20 MC, the 50 MC, the polar arc (PA) \citep{bally88} (blue rectangle), and the molecular loop (ML, pink ellipse) discussed in paper I, are shown. A scale bar of 15 pc (=0.11$\degr$) is displayed in the top right corners.
}
\label{fig-mom0-197kms}
\end{center}
\end{figure}

In Figure~\ref{fig-mom0-197kms} we show the CS($J=2-1$) and CS($J=1-0$) \citep{tsuboi99} integrated intensities.
We overlay the CS($J=2-1$) channel maps on the VLA archival 20-cm radio continuum emission in 
Figure~\ref{fig-chan-outflow}.
The noise levels of these two data sets are comparable. The integrated intensity maps of both transitions are smoothed to 30$\arcsec$ ($\sim$1 pc) for comparison. In general, the structures are consistent in both transitions but the new CS($J=2-1$) line map reveals more details and extended emission than the CS($J=1-0$) line. This is due to the high excitation properties of  the GC \citep[e.g.][]{morris96,maria}.
Two molecular clouds, called 20 km s$^{-1}$ cloud (hereafter 20 MC) and 50 km s$^{-1}$ cloud (hereafter 50 MC), are well studied in the GC  \citep[e.g.,][]{guesten81,tsuboi09,liu12}. These two clouds are part of the central molecular zone (CMZ) \citep[e.g.,][]{morris96}. The 50 MC is known to interact with the SNR SgrA East \citep[e.g.,][]{serabyn92}, and the 20 MC is feeding the circumnuclear disk (CND) \citep[e.g.,][]{ho85}, which is a ring-like feature in between the 20/50 MC
\citep[e.g.,][]{guesten87,jackson93,amo11,harris,mezger,etx,lau,wright,maria,martin12,herrnstein02,
herrnstein05,chris,great,mills13b}. The CND shows a fast rotation  from $-120$ km s$^{-1}$ to 120 km s$^{-1}$. It is known to have a high-excitation state and is, thus, more clearly seen in CS($J=2-1$) than in CS($J=1-0$) \citep[e.g.,][]{mcgary01,mcgary02,maria}. The polar arc (PA) \citep{bally88} is a high-velocity cloud with velocities up to $\ge100$ km s$^{-1}$. It shows a high gas density \citep{tsuboi99} but its origin is not clear. In paper I, we reported that the PA appears to be physically connected to the Galactic disk and it may originate from the disk.  The PA is also the molecular counterpart of the eastern spur of the Galactic Center Lobe (GCL) \citep[e.g.,][]{sofue96}.
The CS($J=2-1$) line emission located at the base of the PA appears both at the negative velocity of $-50$ km s$^{-1}$ and at the positive velocities of 60 to 80 km s$^{-1}$. This double-peak feature is called the molecular loop (ML) in paper I.

In Figure~\ref{fig-chan-outflow}, we note that the CS($J=2-1$) emission seems to have an interesting spatial association with the 20-cm radio continuum. In particular, the molecular gas around $-20$ km s$^{-1}$ to 20 km s$^{-1}$ shows an hourglass-shaped feature (hereafter HG-feature), with a northwest-southeast orientation perpendicular to the Galactic disk, with the openings or cavities that surround the radio halo \citep{pedlar89,zhao14}.
In the negative Galactic plane, the 20/50 MC seem to surround the southern radio halo. 
Above SgrA*, there is also some molecular gas which coincides with the filaments inside the radio lobe.  
We compare this with the previous CS($J=1-0$) line map \citep{tsuboi99}.
To avoid any possible contamination from the 20/50 MC,  we show in Figure~\ref{fig-mom0-outflow} the intensity map integrated for $\pm10$ km s$^{-1}$ based on higher-resolution channel maps (2.5 km s$^{-1}$ resolution). We  overlay the CS line maps on the 20-cm continuum map to show more clearly the HG-feature structures. In general, CS($J=2-1$) and CS($J=1-0$) have a similar morphology. With a comparable sensitivity, our new CS($J=2-1$) line emission shows a more extended structure than CS($J=1-0$).

In Figure~\ref{fig-rgb}, we display the color-composite map of the 20-cm \citep{yusef04} and the MSX E-band 21-$\micron$ map \citep{price01,simpson99,simpson07}. We compare the CS($J=2-1$) line emission with the dust (21 $\micron$) and the emission from free electrons (20 cm). The map shows that the low-velocity gas roughly surrounds the infrared/radio features. The 21-$\micron$ emission shows the prominent Galactic center bubble \citep{simpson07} (GC bubble, or arc bubble in \citealt{ponti15}) near the SgrA radio halo/arc. The northern part of the GC bubble has a radio feature known as the ``radio arc'' \citep[e.g.][]{serabyn87,lang99,yusef84}.
The GC bubble could be produced by the starbursts in the past \citep{simpson07,sofue03}.
Our CS($J=2-1$) line map spatially coincides with the infrared/radio features.
The HG-feature seems to surround the warm dust and the free-electron emission.

In summary, our new high-resolution/excitation CS($J=2-1$) line map suggests that the motions in the central 30 pc are not simply rotation but are also associated with structures in the vertical direction of the Galactic plane. In the following, we will revisit the association of the known features with the HG-feature and the PA.

\subsection{Latitude-Velocity Diagrams}

Latitude-velocity diagrams ($b-v$ diagram) of the CS line data are presented to investigate the extraplanar kinematics.
Figure~\ref{fig-guide} specifies the range for the $b-v$ diagrams.
The $b-v$ diagrams are presented for both CS($J=2-1$) and CS($J=1-0$) data  at a resolution of 30$\arcsec\times$5 km s$^{-1}$ (Figure~\ref{fig-bv-p1-5} to Figure~\ref{fig-bv5}).
In order to improve the signal-to-noise ratios, the $b-v$ diagrams are averaged every
2.6$\arcmin$ in longitude ($\Delta l$) from $l=0.07\degr$ to $l=-0.18\degr$ from Figure~\ref{fig-bv1} to Figure~\ref{fig-bv5}. The gaps between regions are 30$\arcsec$, which corresponds to the convolved beam size.
As shown in Figure~\ref{fig-guide}, region A and region B present the eastern edge of the HG-feature.
Region C covers SgrA* and the CND. Region D and Region E cover the western edge
of the HG-feature.
From a visual inspection, there are several apparent features in the $b-v$ diagrams.
In Figure~\ref{fig-bv-p1-5}, we also present the $b-v$ diagram averaged over the
entire HG-feature (from $l=0.07\degr$ to $l=-0.18\degr$) (Region A to E).  
In the following, we briefly summarize the known/new features with our new wide-field data in this complicated region.

\begin{enumerate}
  \item {\bf The 20 MC and the 50 MC: }
  In Figure~\ref{fig-bv-p1-5}, the 20/50 MC appear south of the Galactic plane from $b=-0.02\degr$ to $b=-0.14\degr$ with velocities from $\sim80$ km s$^{-1}$ to $\sim-30$ km s$^{-1}$. These two molecular clouds are well studied and are known to be located in the Galactic disk \citep[e.g.,][]{guesten81,coil00,mcgary01,serabyn92,liu12}. The morphology of these two clouds is consistent in both CS lines.

  The 50 MC appears in the regions A, B, and C (Figure~\ref{fig-bv1},~\ref{fig-bv2},~\ref{fig-bv3}). \citet{tsuboi09} discussed the expanding properties of the central 50 MC (Region B). A SNR candidate is located in the expanding shell in the 50 MC \citep{tsuboi09}.  The 20 MC appears in the regions D and E. (Figure~\ref{fig-bv4},~\ref{fig-bv5}). A SNR candidate, G359.92-0.09 (wisp; \citealt{ho85}, $l=359.89\degr, b=-0.086\degr$), in the Galactic center region was studied by the ASCA and the Chandra telescopes \citep{senda03}. The interaction of the wisp and the 20 MC was also investigated by \citet{ho85} and \citet{coil00}.

  \item {\bf The Circumnuclear Disk (CND): }
  The CND appears in the regions C and D (Figure~\ref{fig-bv3},~\ref{fig-bv4}) in CS($J=2-1$). As shown in the channel maps. it has high velocities of up to $\pm$120 km s$^{-1}$ and a steep velocity gradient as compared to the Galactic disk. The CND is known to have high excitation \citep{mills13b,maria}.
This explains why the CND appears more significant in CS($J=2-1$) than in CS($J=1-0$)

  \item {\bf The High-Velocity Compact Cloud (HVCC) CO 0.02-0.02:}
 In region B, there is a high-velocity clump ($b=-0.015\degr$; $v\sim$90 km s$^{-1}$) with a broad linewidth $\ge100$ km s$^{-1}$ and with sizes of $\sim3-4$ pc. This is one of the
HVCCs studied by \citet{oka99,oka08} (CO 0.02-0.02). This HVCC is located in the ``finger-like'' ridge that connects to the CND.
This HVCC was suggested to be impacted upon by the SN explosions resulting in an unusually high density and temperature \citep{oka99,oka08, oka11asp}.

  \item {\bf The Polar Arc (PA): }
The PA represents a pair of arcs with positive and negative components above the Galactic plane (green dashed lines, Figure~\ref{fig-bv-p1-5} to Figure~\ref{fig-bv5}) that extend down to $b\le0.23\degr$ and become confused with the Galactic disk. A velocity gradient of $\sim5.5$ km s$^{-1}$ pc $^{-1}$ is seen across the PA. The ML mentioned in Sect.~\ref{cs21-line-map} is located at the contact point of the PA and the Galactic disk around $b=0.03\degr$. Velocities of the ML at the intensity peaks are $\sim60$ km s$^{-1}$ and $\sim-50$ km s$^{-1}$. The ML and the PA appear from region A to region E with a spatial length of $\sim$34 pc. The ML seems to be part of the PA in the $b-v$ diagrams based on their coherent locations and kinematics.
The CS($J=4-3$) CR studied in paper I fills the gap between the positive-velocity knot of the ML and the Galactic disk.
In addition, the positive-velocity component is brighter than the negative-velocity component by a factor of $\sim1.5$.
The $b-v$ diagrams of the PA and the ML indicate an accelerating and expanding motion perpendicular to the Galactic disk.
Similar kinematic structures are seen in outflows from starburst galaxies \citep[e.g.,][]{m82b,tsai09,bolatto13} but on kilo-pc scales.
It is particularly important to note that all the emission north of the Galactic plane, with positive and negative velocities, converges in the $b-v$ diagrams towards SgrA* in the nucleus of the Galaxy in our high-resolution map.

  \item {\bf The Newly Found Bubble?:}
In Figure~\ref{fig-bv-p1-5}, the low-level emission as indicated by the red ellipse in the southern Galactic plane shows a half-curve feature with a central cavity (here, we call this feature B1).
The newly found B1 appears in the regions A, B, C and seems to extend to the northern Galactic plane marked by the blue ellipse, called B2 in Figure~\ref{fig-bv1}.
The kinematics of the B1 can be explained by an expanding bubble in the $b-v$ diagrams. 
Above $b=-0.1\degr$, the B1 is blended with the 50  MC and the $-13$ km s$^{-1}$ cloud.
The structure inside the Galactic disk is very complicated, and it is not clear how to identify the boundary of the northern side of either B1 or B2. For a visual inspection, the center and the radius of B1 are $(b,v)=(-0.098\degr,19~\rm {km~s^{-1}})$ and $(\Delta b, \Delta v)=(0.1\degr,40~\rm {km~s^{-1}})$, respectively. From Figure~\ref{fig-bv1} to Figure~\ref{fig-bv3}, some gas in the 50 MC shows a smooth connection to the B1 from 40 km s$^{-1}$ to 70 km s$^{-1}$.
To avoid any contamination from the CND, we present intensity ratio maps by averaging region A and B (Figure~\ref{fig-bv-ratio2}).
We exclude ratios lower than 3$\sigma$ and hence, the emission in the southern Galactic plane is discarded due to its low-level emission.
High ratios $\ge3.5$ appear in the HVCC CO0.02-0.02 and also in the B2.
At the contact point of the 50 MC and the B1, the ratios are also $\ge3.5$. In general, the remaining disk emission has ratios lower than 2.

  \item {\bf The Hourglass-Shaped (HG) Feature:}
We have identified the HG-feature in the low-velocity channels (Figure~\ref{fig-chan-outflow}). Here, we investigate the kinematics of this structure in the $b-v$ diagrams. 
We are specifically examining the extraplanar nature of this feature. The low-velocity structure of the HG-feature is labeled as yellow box from Figure~\ref{fig-bv1} to Figure~\ref{fig-bv5}, where the regions A/B and D/E trace the eastern and western edges, respectively. The eastern edge of the HG-feature (Figure~\ref{fig-bv1}) corresponds to the blueshifted side of the B1. This side is not blended with the 50 MC and 20 MC, and hence can be seen as a half-shell feature.
The western edge  of the HG-feature shows a half-arc feature and seems to smoothly connect to the 20 MC out to the northern plane (western HG) with increasing velocity. The western HG is mostly located between $-10$ km s$^{-1}$ and 10 km s$^{-1}$.
We also present the intensity ratio map generated by averaging region D and E in Figure~\ref{fig-bv-ratio2}. The high-ratio  ($\ge 3$) data concentrate on the western HG.

\end{enumerate}

\section{Polar Arc: Galactic Center Molecular Outflow?}

In the $b-v$ diagrams (Figure~\ref{fig-bv-p1-5}), we find that the velocity of the PA is linearly increasing  along the Galactic latitude to more than 100 km s$^{-1}$. This can be interpreted as an expanding motion.
Similar features caused by kpc-scale molecular outflows are seen in the nearby starburst galaxies, e.g., NGC 2146, NGC 3628, NGC 253, M82 \citep{tsai09,bolatto13,m82a,m82b}.
These molecular outflows are suggested to be the late evolved stage of superbubbles \citep{yokoo93}, which are driven by intensive stellar winds and supernovae explosions \citep{veilleux05}.
The velocities of the GC outflow are typical values of molecular outflows seen in external starburst galaxies ($\sim100$ km s$^{-1}$ on average; \citealt{veilleus04}), which are generally lower than the velocities of their ionized-gas outflows.
With the implications of (1) extraplanar structures, and (2) expanding/accelerating motions, we  suggest that the polar arc is associated with the molecular outflow. We then estimate the kinetic properties of the outflow by assuming a constant acceleration. For the positive-velocity ridge, we use a velocity of $\sim100$ km s$^{-1}$ and a length of 34 pc (as shown in Figure~\ref{fig-bv-p1-5}), which then gives an acceleration $a_{\rm pos}$ = $0.5\times(100~\rm km~s^{-1})^{2}/(34~pc)$ = 4.8$\times10^{-12}$ km s$^{-2}$, and a dynamical time scale $t_{\rm pos}$ = (100 km s$^{-1}$)/4.8$\times10^{-12}$ km s$^{-2}$ = 6.7$\times10^{5}$ years.  We note that this value is averaged over the structure. As shown in paper I, the PA belongs to the eastern protrusion of the GCL (see Figure 20 in paper I). The time scale of the PA is close to the dynamical time scale of the GCL (1 Myr) \citep{bland03}.
Implications of the extraplanar PA can be traced back to the disk region (the ML), providing clues of the launching point of the PA, and perhaps a link to the GCL at parsec scale.
At kpc-scale, fossil imprints of past explosions in the GC were reported in \citet[e.g.,][]{bland03,su10,crocker11}. For example, there is evidence that the GC experienced episodic starbursts a few Myr ago, producing more than 100 SNe during the entire starburst
\citep[q.v.][]{tamblyn93,Sjouwerman98,simpson99,hartmann94,Ozernoy96,carretti13}. Detailed modeling of the low-excitation fine-structure line spectrum \citep{lutz99} suggests a starburst event about 7 Myr ago. It is also suggested that SgrA* went through a Seyfert phase in the recent past \citep[q.v.][]{zubovas11,guo12,bland13}.
\citet{guo12} propose that the SgrA* jets, which inflated the Fermi bubble \citep{su10}, formed $1-3$ Myr ago and persisted for $0.1-0.5$ Myr. The total energy of the SgrA* jets is in the range of $10^{55-57}$ erg.
The {\it ASCA} $2-10$ keV observations of the CMZ also provide evidence that SgrA* was $10^{5}$ times more active in the past $10^{3}$ years. The fluorescent X-ray emission from cold iron atoms in the molecular clouds of the CMZ is possibly due to the X-ray irradiation from SgrA* flares \citep[q.v.][]{sunyaev93,koyama96,ponti10}. 
\citet{bland13} also suggest that the H$\alpha$ emission of the Magellanic Stream arose from an accretion flare of SgrA* $1-3$ Myr ago. The required star-formation rate to produce the H$\alpha$ emission of the Magellanic stream is at least two orders of magnitude larger than what can be generated by the star formation history of the GC.
In this paper, we have confirmed that the PA originates from the Galactic disk. \citet{henshaw16} also show that the PA seems to extend from Arm I \citep{sofue95}, but with an offset in velocity. Our results suggest that in the central 30 pc of the GC, the feedback from past explosions could alter the orbital path of the molecular gas down to the central tenth of parsec.

\section{Extraplanar Features in the Galactic Center}

In the $b-v$ diagrams (Figure~\ref{fig-bv-p1-5}), the molecular feature B1 appearing below $b=-0.1\degr$ suggests an expanding bubble.
The  B1  has higher CS($J=2-1$)/CS($J=1-0$) line ratios ($\sim$3) than the ambient 20/50 MC by a factor of 2. The B1 has a counterpart in SiO($J=2-1$) \citep{tsuboi11} suggesting that the low-level emission of B1 is shocked.
At $v=-20$ to $v=60$ km s$^{-1}$, the B1 seems to interact with the 20/50 MC, as these clouds curve around the entire southern part of the shell in the $b-v$ diagrams (marked by the ellipse of B1). The northern boundary ($b\ge-0.1\degr$) of the  B1 is not clear due to the complicated structures inside the 20/50 MC. It is possible that the B1 extends to the 20/50 MC as hinted in the ratio map, marked as B2 (Figure~\ref{fig-bv-ratio2}). The B2 has ratios higher than 3 and shows a shell-like feature in the 20/50 MC.
In region A (Figure~\ref{fig-bv1}), above $b=0\degr$, the ''low-velocity'' emission tracing the eastern edge of the HG-feature (yellow box) is hard to identify because of the confusion of the Galactic disk. 
This ''low-velocity'' emission might also belong to the ``high-velocity'' part of the B1 if we consider that the B1 extends up to $b=0.05\degr$.
In this sense, the eastern edge of the HG-shaped feature might be part of the B1.
In any case, we estimate the lower limit of the kinetic parameters
with the B1. With an expanding velocity of $\sim40$ km~s$^{-1}$, the kinetic energy of the northern  HG-feature is $\frac{1}{2}M_{\rm shell}V_{\rm shell}^{2}$, which is at the order of $\sim10^{49}$ erg with an H$_2$ mass of $\sim1000~{M}_{\odot}$. The H$_2$ mass is adopted from a Local Thermal Equilibrium (LTE) condition, with an excitation temperature of 20 K (paper I), a CS abundance of $5\times10^{-9}$ \citep{martin08}, and the measured brightness temperature from the CS($J=1-0$) line. The shell kinetic energy could be different by a factor of 0.1 to 10 for different CS abundances and excitation temperatures.
The expansion time of the B1 is $\sim3\times10^{5}$ years with the above expansion velocity of 40 km s$^{-1}$ and a size of 13 pc. We notice that the time scales of the  B1 and the PA are similar.
The time scale is larger than the age of SgrA East (a few 10$^{4}$ yr; e.g., \citealt{herrnstein05}). However, we note that the current measured expansion velocity sets an upper limit on the age, because the expansion speed was most likely larger in the past.   
Nevertheless, the identification of these individual shells is not unique or certain.  There is also the possibility that all of the features are related in that they may be complexities on top of a general outflowing shell.

The western edge of the HG-feature (western-HG) is located in the region D (Figure~\ref{fig-bv4}). There is a velocity continuation of the  20 MC above SgrA*. This continuation traces the western edge of the HG-feature above the Galactic plane from $(b,v)=(-0.03\degr,-10~{\rm km~s^{-1}})$ to $(b,v)=(0.06\degr,30~{\rm km~s^{-1}})$. The velocity of this continuation is increasing out to the Galactic latitude and corresponds to the western-HG. Rather than a closed expanding bubble, this feature seems to be entrained and swept into the HG-feature from the 20 MC, which shows a similar kinematic behavior as the PA. The ratios ($\ge4.5$) of the western-HG are higher than those of the cold disk emission ($\le2.5$), which suggests a higher excitation state in the western-HG. However, the structures in this region are complicated, and it is difficult to identify structures in the $b-v$ diagrams.
Moreover, the current results do not show evidence that the HG-feature is a physical bipolar feature because the western and eastern edges do not show a coherent shell expansion. 
The asymmetry of the structures in the $b-v$ diagrams might suggest that the outflow phenomenon must be greatly influenced by the asymmetric circumnuclear environment.

In the following, we discuss the physical association between the HG-feature and the radio halo/lobe. The bipolar radio lobe has counterparts in X-ray emission \citep{maeda02,markoff10}. Mixtures of the thermal S and Si emission in the radio lobe suggest a thermal origin of the hot plasma \citep{park04}, heated by the shocks from SNe and massive stellar winds. The interaction of the hot plasma with the ambient dense cloud might also produce the observed neutral Fe line. 
It is unclear whether the hot plasma from the radio lobe in situ can entrain the molecular gas.
We  do a self-consistent check of whether the ionized gas of the bipolar radio-lobe is able to entrain the molecular gas from the disk clouds, assuming it is moving outward perpendicular to the GC. For the given emission measure of 2.7$\times10^{5}$ pc cm$^{-6}$ and the source size of 4$\arcmin$ \citep{pedlar89}, we derive the electron density $n_{\rm e}$ to be 7.8$\times10^{4}$ cm$^{-3}$. The electron temperature is $T_{\rm e}$ = 5000 K adapted from \citet{pedlar89}. Therefore, we estimate the thermal ionized gas pressure to be $P_{\rm i}/k=n_{\rm e}T_{\rm e}\sim4\times10^{8}$ cm$^{-3}$ K. 
The thermal pressure  of the molecular gas is derived from $P_{\rm m}/k=n_{\rm H_2}T_{\rm kin}$, where $P_{\rm m}/k$ is the gas thermal pressure, $n_{\rm H_2}$ is the number density  of the molecular gas, and $T_{\rm kin}$ is the kinetic temperature. We assume $n_{\rm H_2}$ is $10^{5}$ cm$^{-3}$ (the critical density of CS($J=2-1$)), and a kinetic temperature of 20 K (paper I). The $P_{\rm m}/k$ is then derived as 2$\times10^{6}$ cm$^{-3}$ K ($P_{\rm m}=7\times10^{-10}$ erg cm$^{-3}$). The thermal pressure of the ionized lobe is, thus, sufficient to entrain the molecular gas. An alternative way is to measure the momentum of the radio halo and compare it with the molecular gas \citep[e.g.,][]{matsu04,matsu07}. However, we are currently lacking velocity information of the radio halo. Future kinematic measurements of the radio halo will help to further elucidate the interaction between the ionized and molecular gas.

\section{SUMMARY}

From our 1-pc resolution CS($J=2-1$) line maps, we find that the molecular gas in the central 30 pc of the Galactic Center has a morphology that is reflecting recent explosions $(3-6)\times10^{5}$ years ago. As a follow-up study of paper I, we revisit the idea that the PA is a molecular outflow launched from the Galactic disk. The linearly increasing velocity of the PA suggests that it is moving out of the Galactic disk.
Near the systematic velocity, the molecular gas shows an HG-feature, which might be a mixture of multiple explosive events. The north-western edge of the HG-feature might trace the entrainment from the 20 MC from $b=-0.07\degr$ to $b=0.06\degr$. The southern part of the eastern edge of the HG-feature suggests an expanding bubble (B1). The low-level emission of B1 south to the Galactic plane is shocked. These individual features might be complexities on top of a general outflowing shell.

\acknowledgements
We thank the reviewer for a thoughtful review and suggestions to improve our manuscript.
We thank Dr. Yusef-Zadeh for providing us with the VLA 20 cm continuum map. Hsieh, Pei-Ying is supported by the National Science Council (NSC) and the Ministry of Science and Technology (MoST) of Taiwan through the grants NSC 100-2112-M-001-006-MY3, NSC 97-2112-M-001-021-MY3, and MoST 103-2112-M-001-032-MY3.

This research made use of data products from the Midcourse Space Experiment. This research has also made use of the NASA/ IPAC Infrared Science Archive, which is operated by the Jet Propulsion Laboratory, California Institute of Technology, under contract with the National Aeronautics and Space Administration.


\begin{figure}[hp]
\begin{center}
\epsscale{0.7}
\includegraphics[angle=0,scale=0.6]{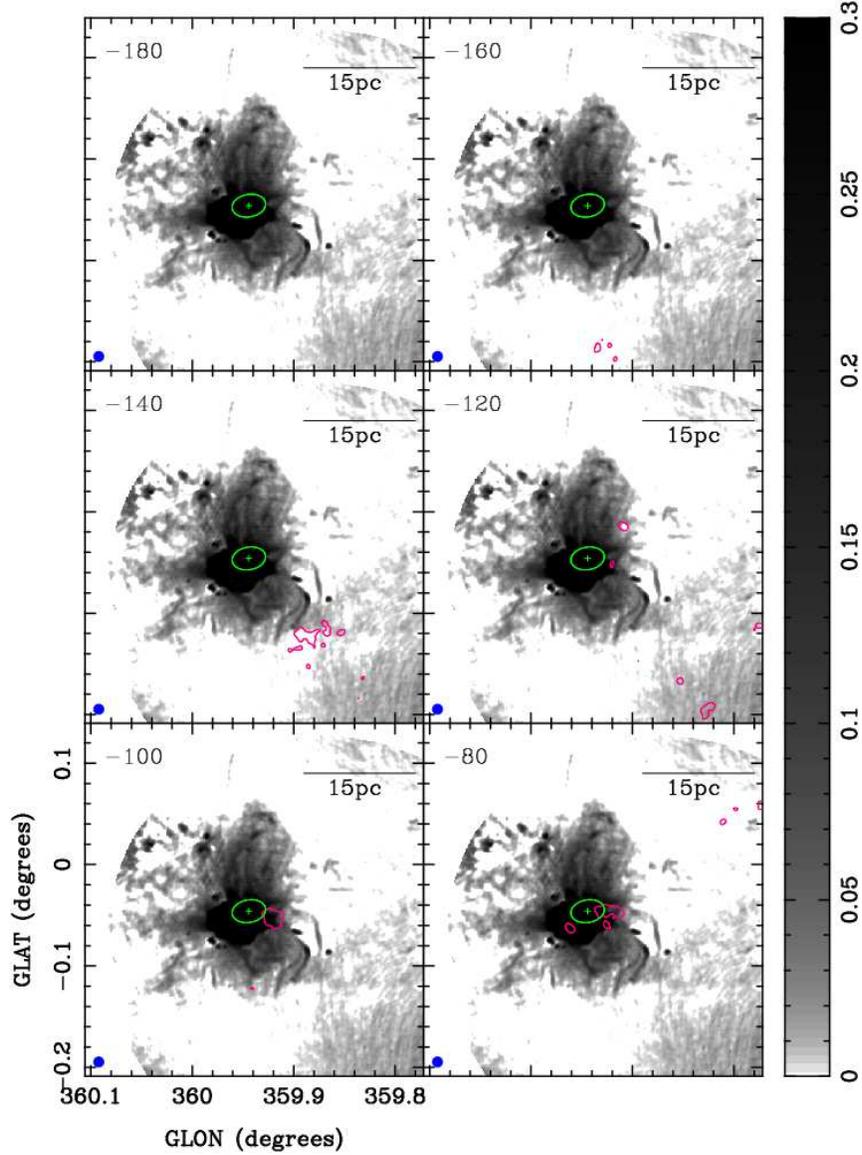}  
\caption[]{CS($J=2-1$) line channel maps (contours) overlaid on archival VLA 20-cm continuum. The (0,0) offset corresponds to $(l,b)=(359.939\degr, -0.056\degr)$. The intrinsic velocity resolution is 2.5 km s$^{-1}$. We show a resolution of 40 km s$^{-1}$ to reduce the number of panels. The beam size of the CS($J=2-1$) line is shown in its original resolution of 21$\arcsec$ (dark blue filled circle in lower left corner). The contour levels are, 5, 10, 20, 30, 40, 50$\sigma$, where $1\sigma=0.04$K ($T_{\rm A}*$). The green ellipse marks the circumnuclear disk (CND). The cross shows the location of SgrA*. The HG-feature discussed in the text is shown with cyan dots. The location of the 20 MC, 50 MC, the HVCC CO0.02-0.20, the PA, and the ML (yellow ellipse) are labeled.
}
\label{fig-chan-outflow}
\end{center}
\end{figure}

\addtocounter{figure}{-1}
\begin{figure}[hp]
\begin{center}
\epsscale{0.5}
\includegraphics[angle=0,scale=0.6]{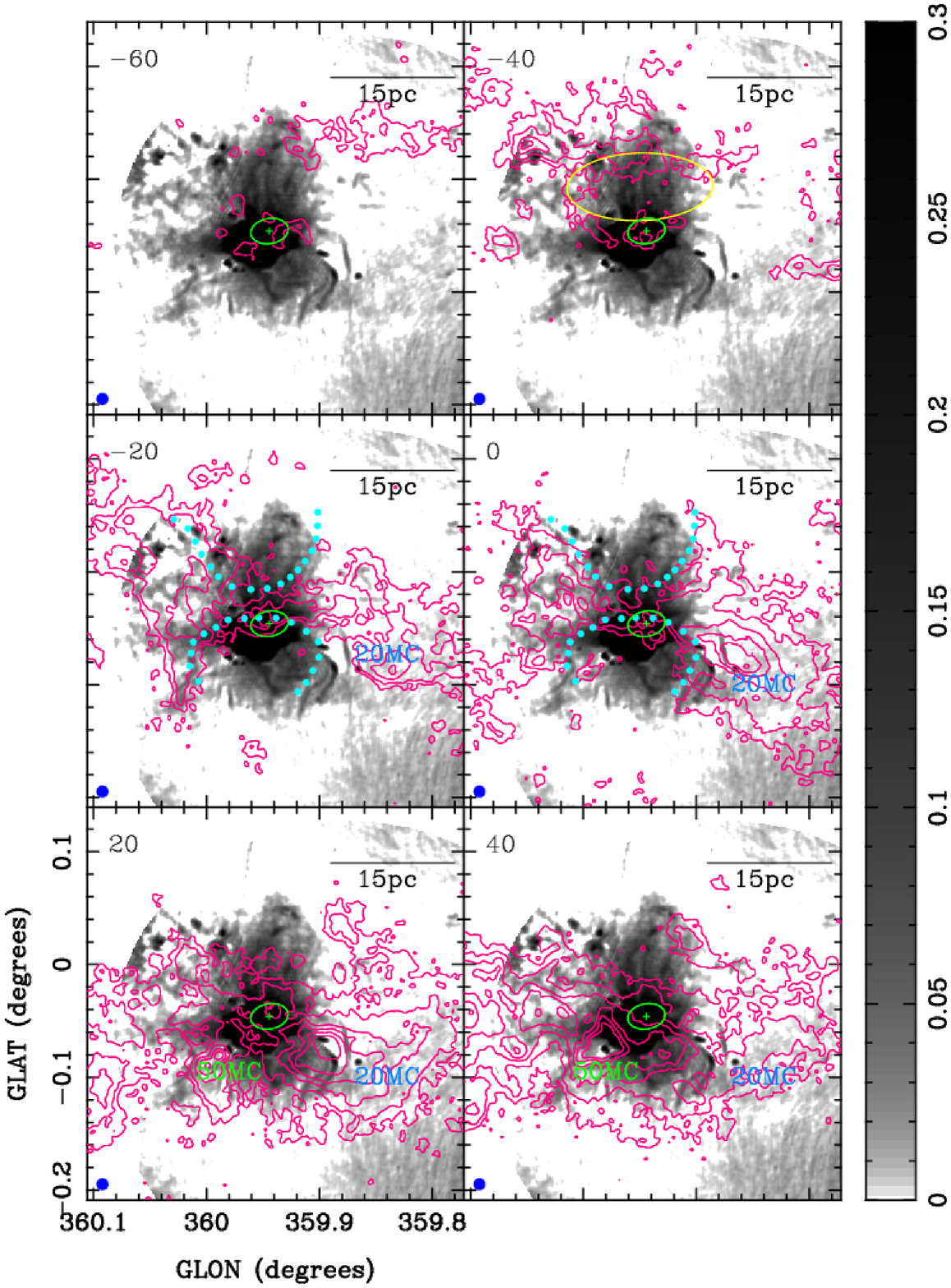}
\caption[CS($J=2-1$) line channel maps (contour) overlaid on archival VLA 20-cm continuum.]{\small Continued.}
\label{}
\end{center}
\end{figure}

\addtocounter{figure}{-1}
\begin{figure}[hp]
\begin{center}
\epsscale{0.5}
\includegraphics[angle=0,scale=0.6]{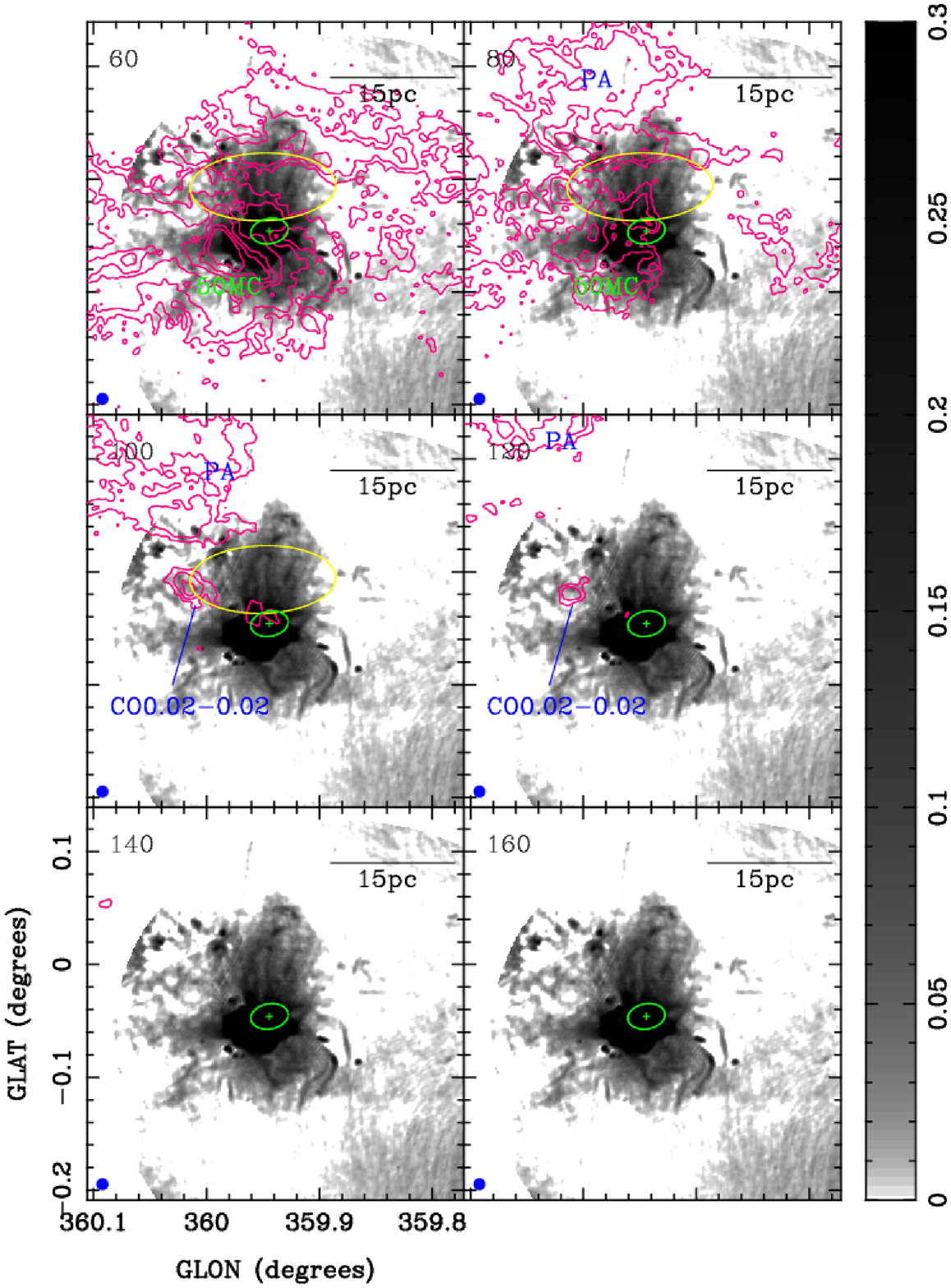}
\caption[CS($J=2-1$) line channel maps (contour) overlaid on the archival VLA 20-cm continuum.]{\small Continued.}
\label{}
\end{center}
\end{figure}

\begin{figure}[hp]
\begin{center}
\epsscale{0.5}
\includegraphics[angle=0,scale=0.45]{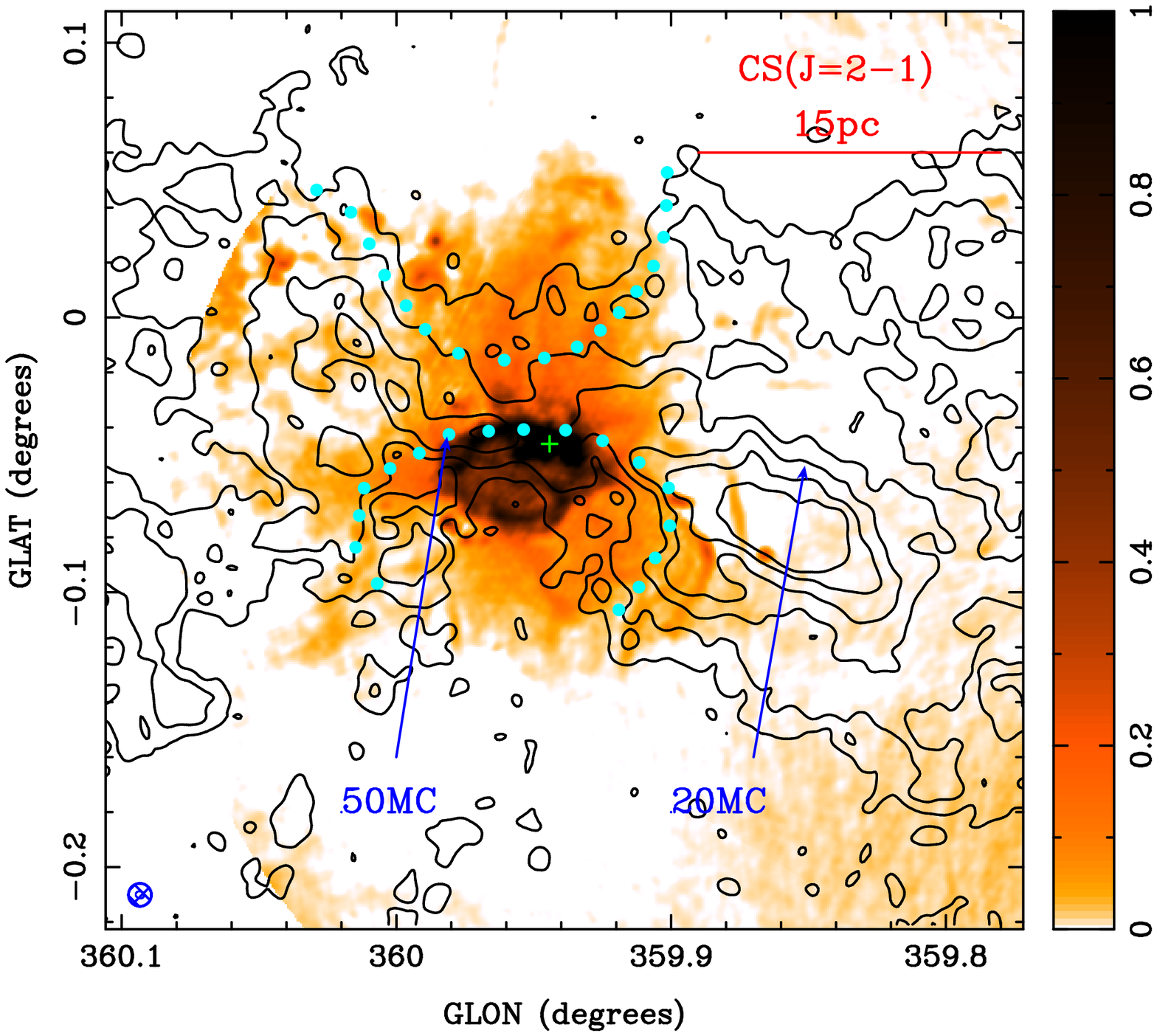}
\includegraphics[angle=0,scale=0.45]{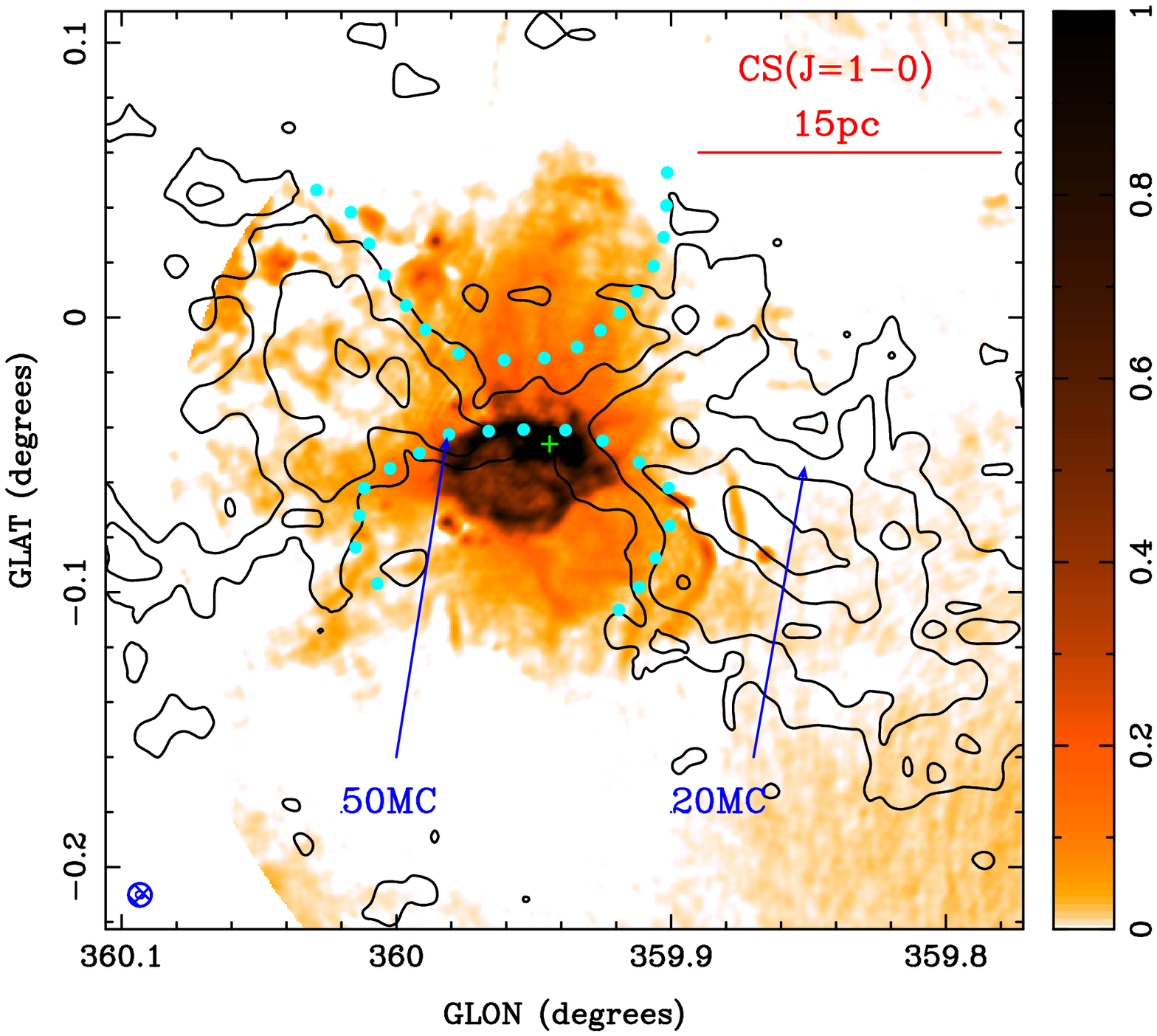}
\caption[]{Top: CS($J=2-1$) line emission (contours) integrated from $-10$ km s$^{-1}$ to 10 km s$^{-1}$ overlaid on  archival VLA 20-cm continuum (color; Jy/beam). Bottom: CS($J=1-0$) line emission (contours) \citep{tsuboi99} integrated from $-10$ km s$^{-1}$ to 10 km s$^{-1}$ overlaid on archival VLA 20-cm continuum emission (color). The contour levels are 5, 10, 20, 30, 40, 50$\times$1.47 K km s$^{-1}$ for both images. The beam size is 30$\arcsec$ for both images (blue circles in lower left corners). The HG-feature is marked with cyan dots. A scale bar of 15 pc (=0.11$\degr$) is shown. 
}
\label{fig-mom0-outflow}
\end{center}
\end{figure}

\begin{center}
\begin{figure}[hp]
\includegraphics[angle=0,scale=0.4]{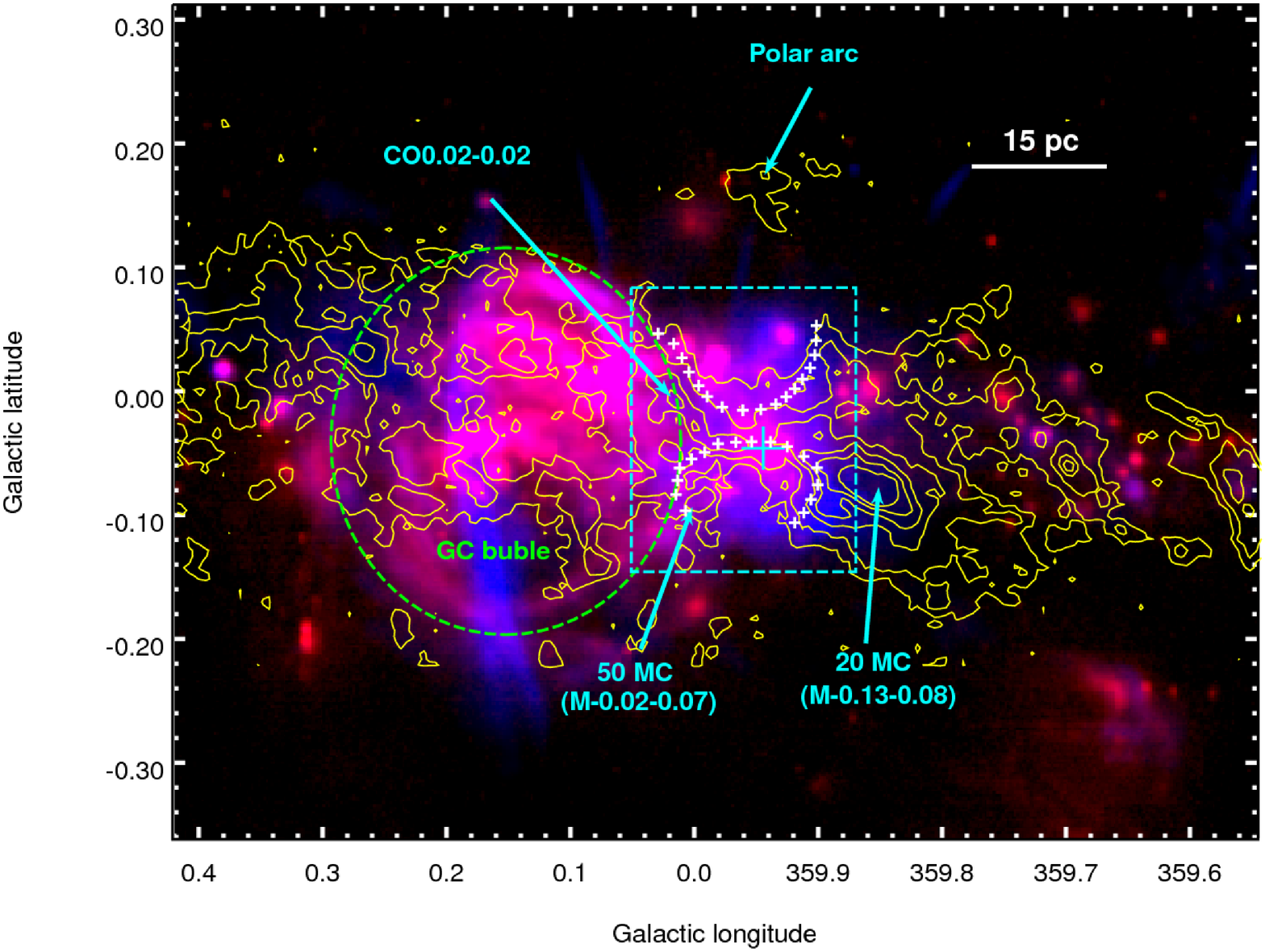}
\caption[]{Color-composite image of 20-cm VLA map (blue) (\citet{yusef04}, courtesy from Dr. Yusef-Zadeh) and archival MSX E-band (21 $\micron$) map (red) \citep{price01}. The CS($J=2-1$) line contour maps integrated over $\pm10$ km s$^{-1}$ are overlaid on it. The hourglass feature is marked with white small crosses and a dashed line box. SgrA* is marked as a large cyan cross. The GC bubble seen in 21 $\micron$ is marked with a green ellipse. A scale bar of 15 pc (=0.11$\degr$) is shown.
}
\label{fig-rgb}
\end{figure}
\end{center}

\begin{figure}[hp]
\begin{center}
\hspace*{-2cm}
\includegraphics[angle=0,scale=0.4]{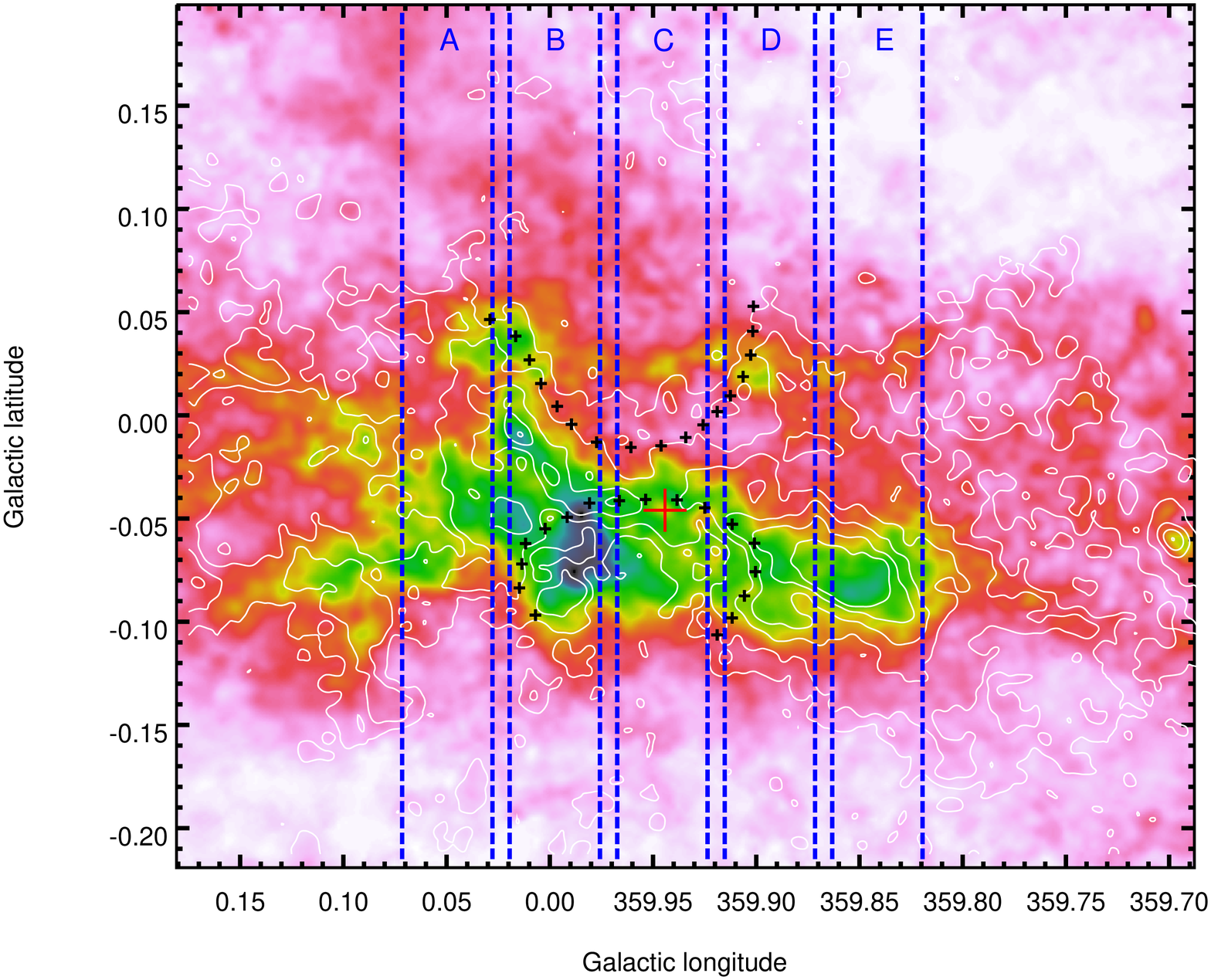}
\caption[]{
Regions made for the latitude velocity diagrams are shown in the CS($J=2-1$) line map. Each region covers a width of $2.6\arcmin$ in longitude. The color map is identical to the one in Figure~\ref{fig-mom0-197kms} and the contour image is identical to Figure~\ref{fig-mom0-outflow}. Regions A, B, C, D, E are marked. The central large red cross marks the position of SgrA*. Small black crosses mark the HG-feature. The spacing between regions corresponds to the convolved beam size of 30$\arcsec$.
}
\label{fig-guide}
\end{center}
\end{figure}

\begin{figure}[hp]
\begin{center}
\epsscale{0.5}
\includegraphics[angle=90,scale=0.65]{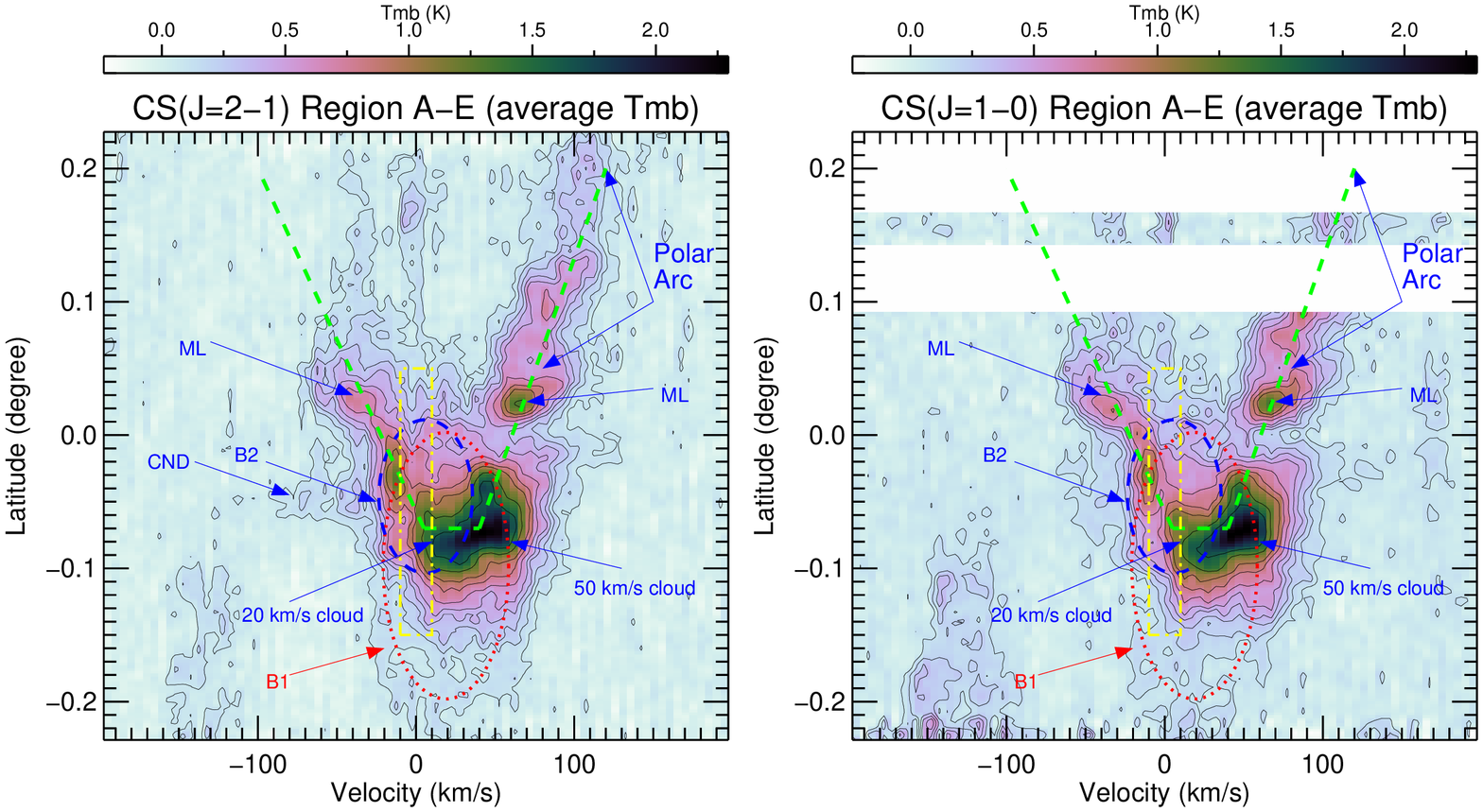}
\caption[]{
Latitude velocity diagrams of region A-E. We use the AIPS task SQASH to collapse longitudinal planes (see Fig.~\ref{fig-guide}) in a cube onto one plane. Diagrams are made by averaging data over the longitude from $l=0.07\degr$ to $-0.18\degr$. The x-axis shows the velocity and the y-axis shows the latitude. The temperature is shown in color and contours in units of brightness temperature ($T_{\rm mb}$). The contours levels are: CS($J=2-1$): 4, 8, 12, 16, 20, 27, 35, 42, 56, 70\% of the peak, where the peak is 2.3 K; CS($J=1-0$): identical levels but the peak is 1.5 K.
The yellow rectangle marks the low-velocity HG-feature. The 50 km s$^{-1}$ and 20 km s$^{-1}$ cloud are located in the Galactic disk. The PA extends out of the Galactic disk to $b\sim0.23\degr$ with increasing velocity and connects to the Galactic disk via the ML (paper I). The PA is possibly a fossil expanding outflow with a time scale of $6\times10^{5}$ years. The B1/B2 labels show the position of an expanding bubble south of the Galactic disk. Other known features of the CND and the HVCC CO0.02-0.02, are labeled as well.
}
\label{fig-bv-p1-5}
\end{center}
\end{figure}

\begin{figure}[hp]
\begin{center}
\epsscale{0.5}
\includegraphics[angle=90,scale=0.65]{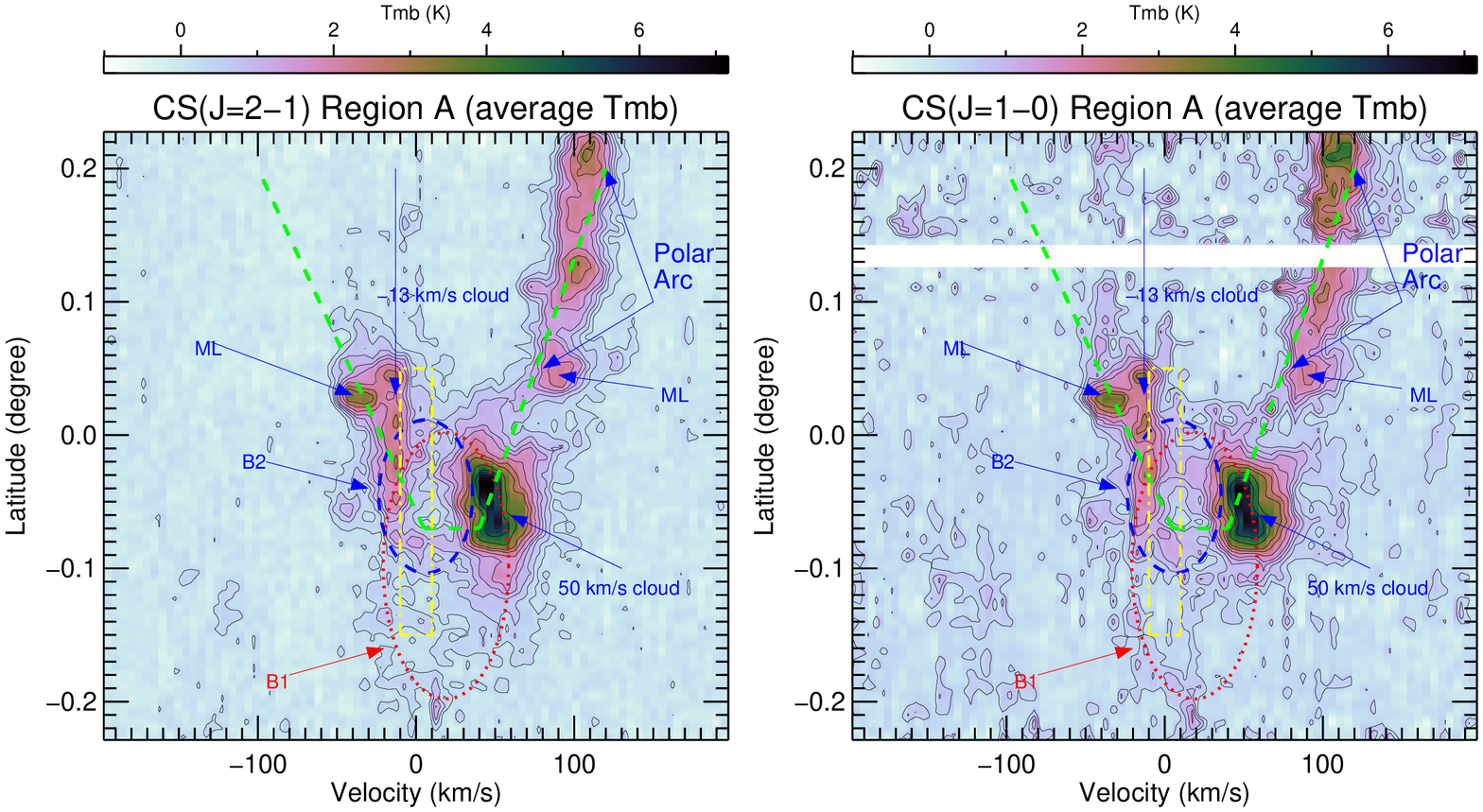}
\caption[]{
Latitude velocity diagrams of region A, which covers the eastern edge of the HG-feature. Identical to Figure~\ref{fig-bv-p1-5} but for region A,  averaging is performed over the longitude from $l=0.07\degr$ to $0.028\degr$. The contour levels are:
CS($J=2-1$): 4, 8, 12, 16, 20, 27, 35, 42, 56,70\% of the peak, where the peak is 7.2 K; 
CS($J=1-0$): identical levels but the peak is 2.2 K.
The yellow rectangle marks the eastern side of the HG-feature. The locations of the features are mentioned in the text. 
}
\label{fig-bv1}
\end{center}
\end{figure}

\begin{figure}[hp]
\begin{center}
\epsscale{0.5}
\includegraphics[angle=90,scale=0.65]{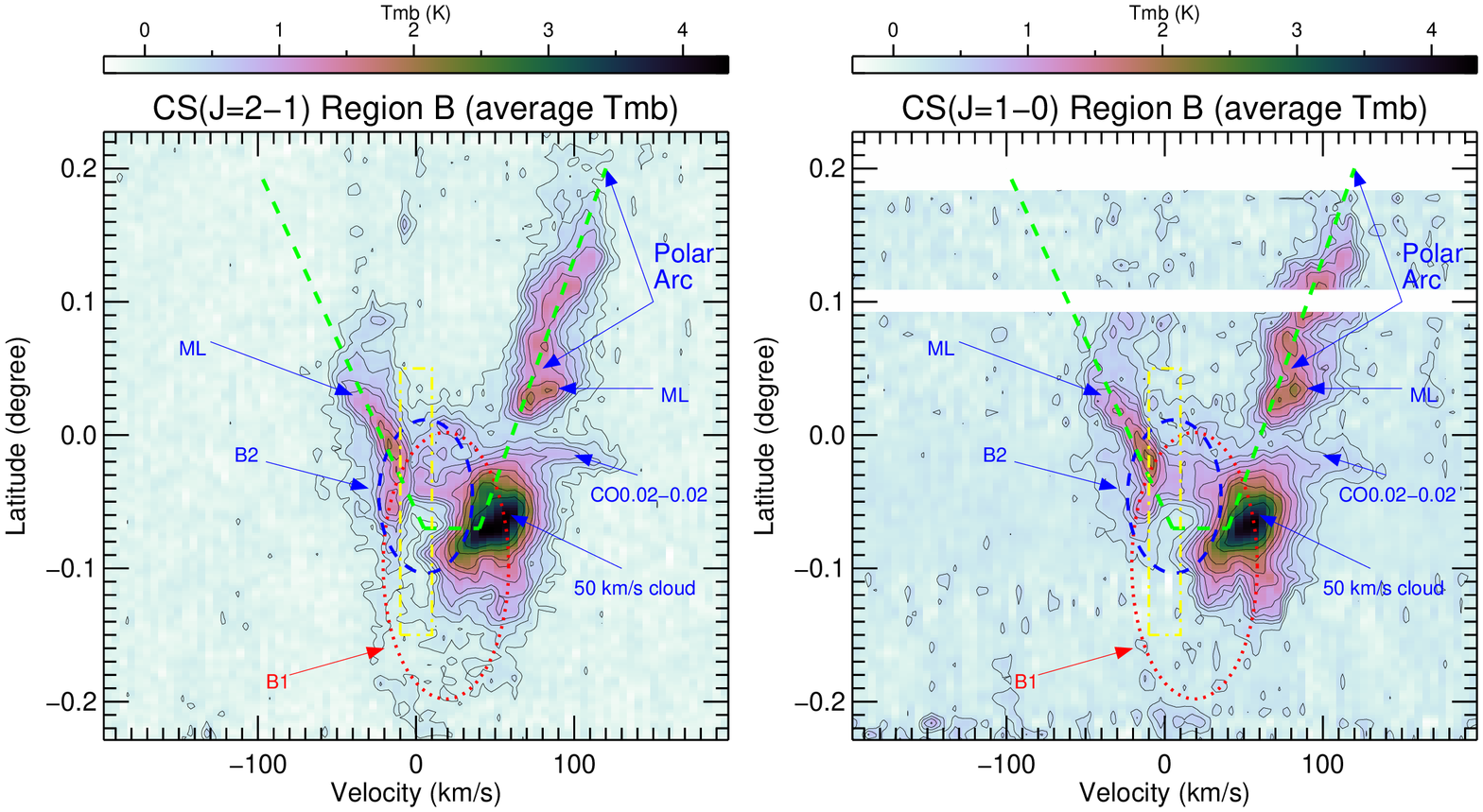}
\caption[]{
Latitude velocity diagrams of region B, which covers the eastern edge of HG-feature. Identical to Figure~\ref{fig-bv-p1-5} but for region B, averaging is performed over the longitude from $l=0.019\degr$  to $359.98\degr$. The contour levels are:
CS($J=2-1$): 4, 8, 12, 16, 20, 27, 35, 42, 56, 70\% of the peak, where the peak is 4.3K;
CS($J=1-0$): identical levels but the peak is 3.0 K.
The yellow rectangle marks the eastern side of the HG-feature.  
}
\label{fig-bv2}
\end{center}
\end{figure}

\begin{figure}[hp]
\begin{center}
\epsscale{0.5}
\includegraphics[angle=90,scale=0.65]{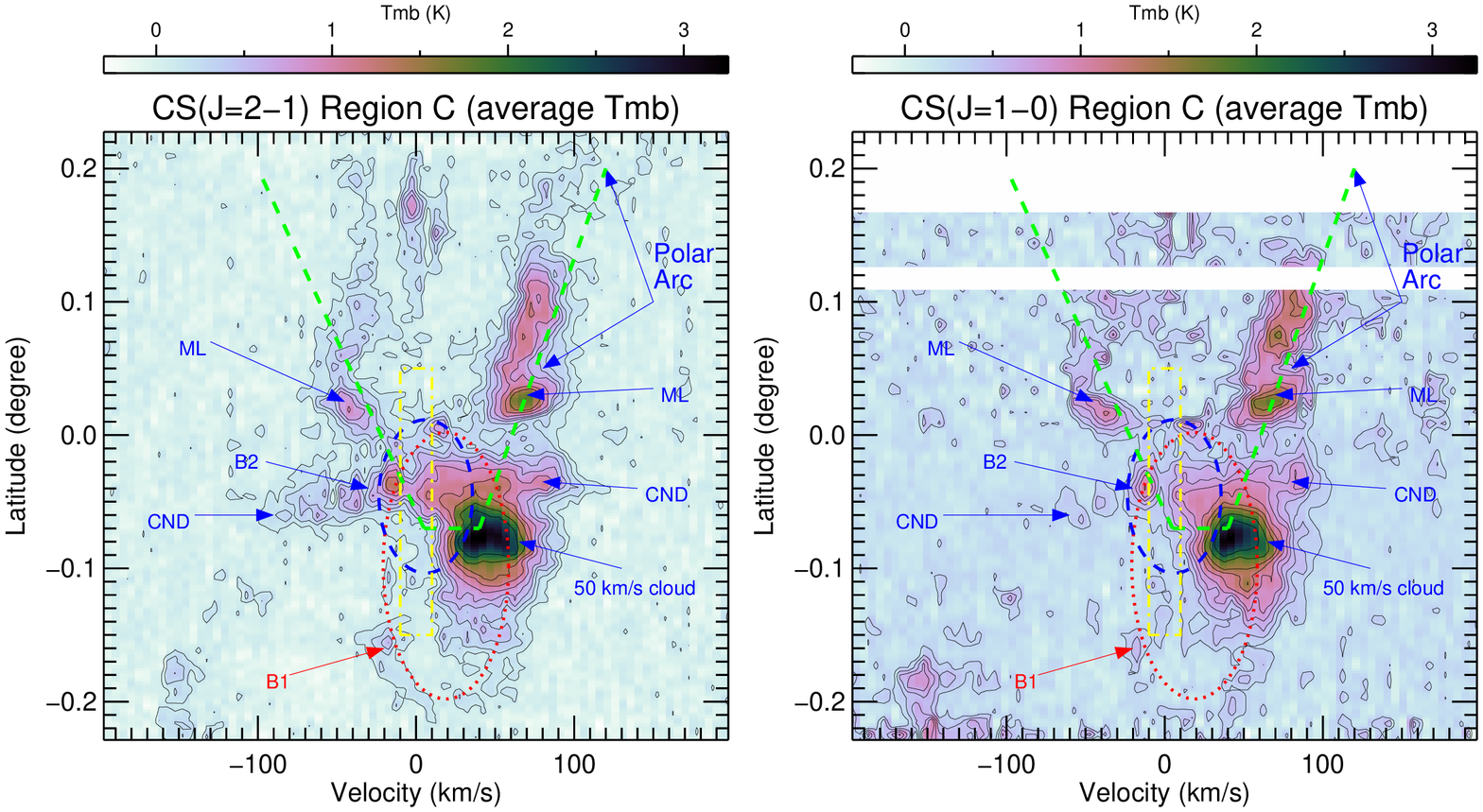}
\caption[]{
Latitude velocity diagrams of region C, which covers SgrA*. Identical to Figure~\ref{fig-bv-p1-5} but for region C, averaging is performed over the longitude from $l=359.97\degr$ $359.92\degr$.
The contour levels are:
CS($J=2-1$): 4, 8, 12, 16, 20, 27, 35, 42, 56, 70\% of the peak, where the peak is 3.3 K;
CS($J=1-0$): identical levels but the peak is 2.3 K. The yellow rectangle marks the center of the HG-feature.
The features are mentioned in the text.
}
\label{fig-bv3}
\end{center}
\end{figure}

\begin{figure}[hp]
\begin{center}
\epsscale{0.5}
\includegraphics[angle=90,scale=0.65]{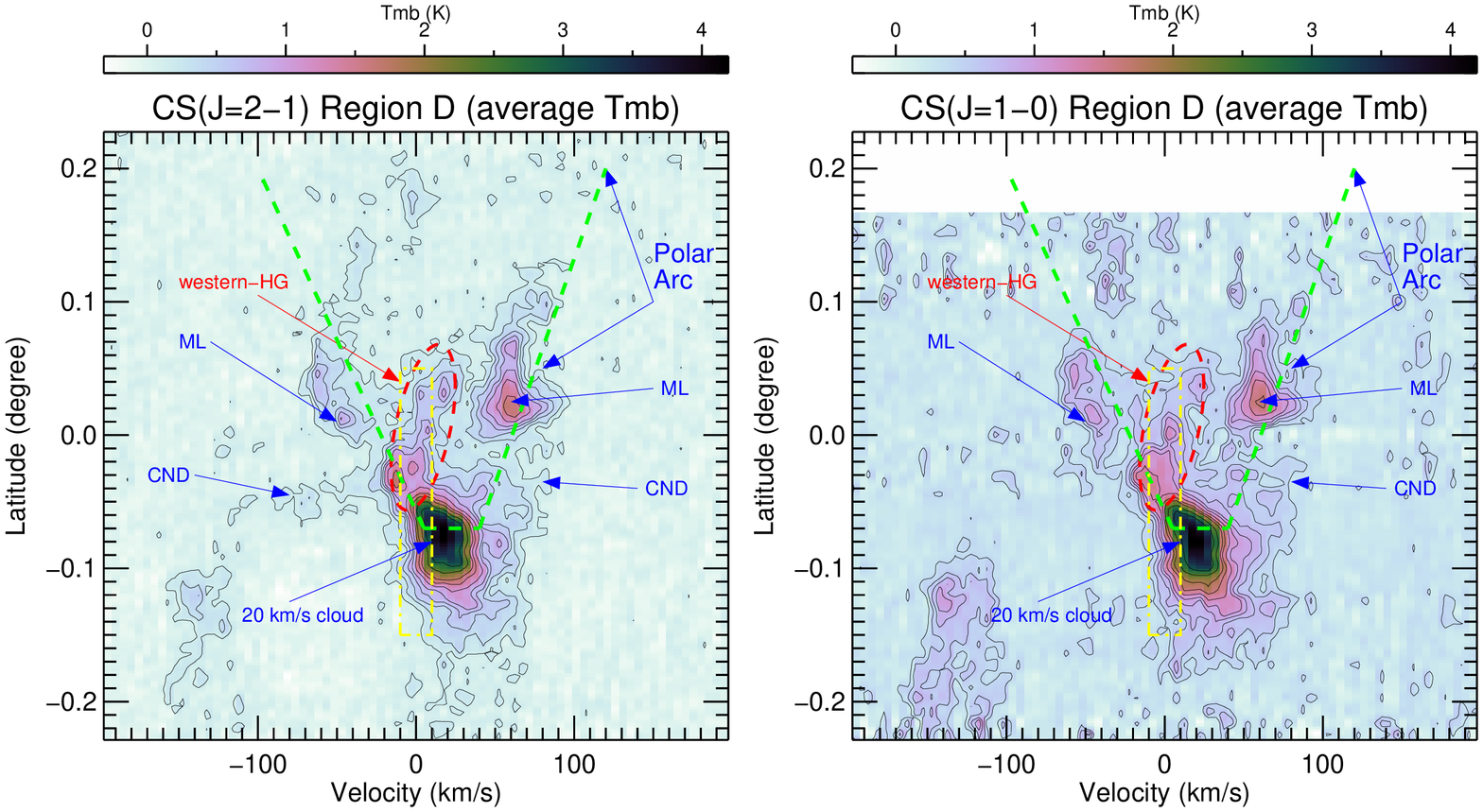}
\caption[]{
Latitude velocity diagrams of region D, which covers the western edge of the HG-feature. Identical to Figure~\ref{fig-bv-p1-5} but for region D, averaging is performed over the longitude from $l=359.92\degr$ $359.87\degr$. The contour levels are:
CS($J=2-1$): 4, 8, 12, 16, 20, 27, 35, 42, 56, 70\% of the peak, where the peak is 4.2 K;
CS($J=1-0$): identical levels but the peak is 2.6 K. The yellow rectangle marks the western side of the HG-feature. The locations of the features are mentioned in the text. 
}
\label{fig-bv4}
\end{center}
\end{figure}

\begin{figure}[hp]
\begin{center}
\epsscale{0.5}
\includegraphics[angle=90,scale=0.65]{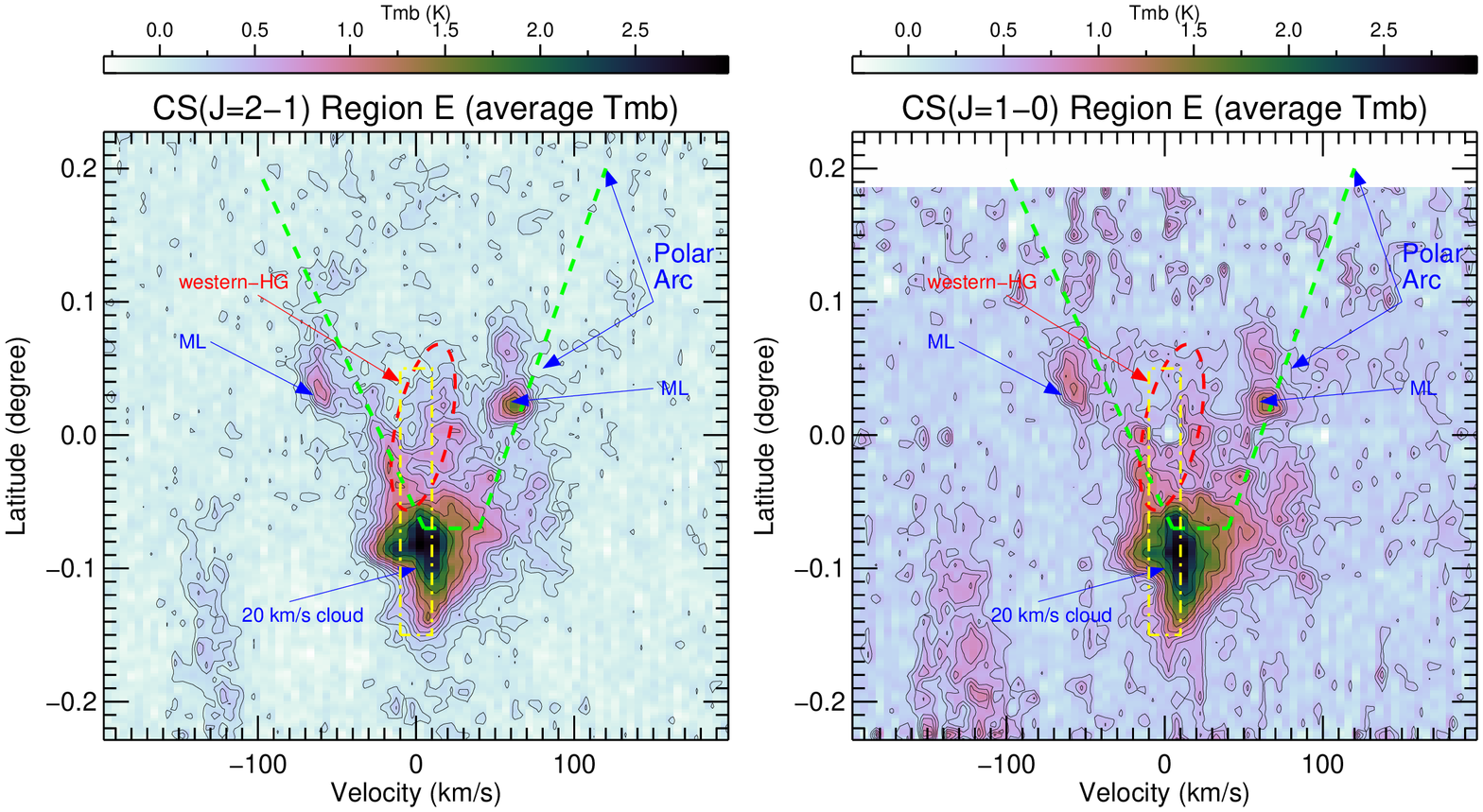}
\caption[]{
Latitude velocity diagrams of region E, which covers the western edge of the HG-feature. Identical to Figure~\ref{fig-bv-p1-5} but for region E, averaging is performed over the longitude from $l=359.86\degr$ $-0.18\degr$. The contour levels are:
CS($J=2-1$): 4, 8, 12, 16, 20, 27, 35, 42, 56, 70\% of the peak, where the peak is 3.0 K;
CS($J=1-0$): identical levels but the peak is 1.8 K.
The yellow rectangle marks the western side of the HG-feature. The locations of the features are mentioned in the text. 
}
\label{fig-bv5}
\end{center}
\end{figure}

\begin{figure}[hp]
\begin{center}
\epsscale{0.5}
\includegraphics[angle=90,scale=0.65]{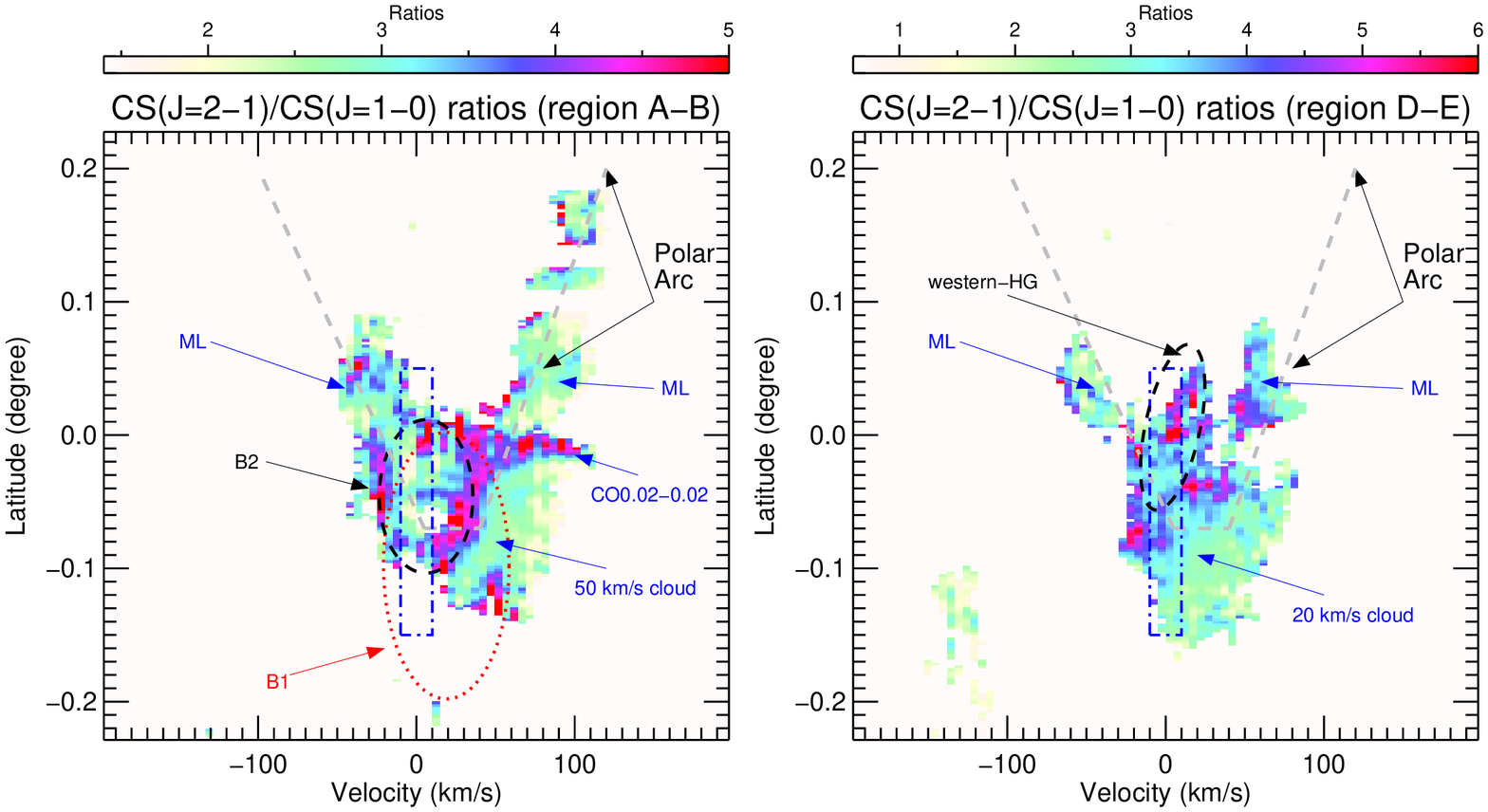}
\caption[]{
Latitude velocity diagrams of the CS($J=2-1$)/CS($J=1-0$) line ratios.
Left: ratios are made by averaging the $b-v$ diagrams with region A and B. Right: ratios are made by averaging the $b-v$ diagrams with region D and E. The locations of the features are mentioned in the text.
The blue rectangle marks the eastern and western side of the HG-feature in the left and right panel, respectively.
}
\label{fig-bv-ratio2}
\end{center}
\end{figure}

\end{document}